\documentclass[fleqn,twoside]{article}
\usepackage{espcrc2}
\usepackage{graphicx}
\usepackage[figuresright]{rotating}

\newcommand{\n}{{\not\hspace{-0.5ex}\nabla}}
\newcommand{\A}{{\not\hspace{-0.5ex}A}}
\newcommand{\N}{{\not\hspace{-0.8ex}D_L}}

\newcommand{\AmS}{{\protect\the\textfont2
  A\kern-.1667em\lower.5ex\hbox{M}\kern-.125emS}}
\title{Vacuum Polarization by a Magnetic Flux in a Cosmic String
Background}

\author{J. Spinelly \address[MCSD]{Departamento de F\'\i sica, Universidade 
        Estadual da Para\'\i ba, \\ 
        Av. Juv\^encio Arruda S/N, Campina Grande, PB, Brazil}%
        \thanks{spinelly@fisica.ufpb.br} and
        E. R. Bezerra de Mello\address[MCSD]{Departamento de F\'\i sica, Universidade Federal da Para\'\i ba \\
        Caixa Postal 5008, 58.059-970, Jo\~ao Pessoa, PB, Brazil}\thanks{emello@fisica.ufpb.br}}
       
\begin{document}

\begin{abstract}
In this paper we analyse the vacuum polarization effects due to a magnetic flux 
on massless fermionic fields in a cosmic string background. Three distinct 
configurations of magnetic fields are considered. In all of them the magnetic fluxes 
are confined in a long cylindrical tube of finite radius.
\vspace{1pc}
\end{abstract}

\maketitle

\section{INTRODUCTION}

The vacuum polarization effects due to a magnetic field confined in a tube 
of finite radius in Minkowski spacetime were first analysed by Serebryanyi 
\cite{Serebryani}. A few years later, Guimar\~aes and Linet \cite{Emilia} and 
Linet \cite{Linet}, calculated these effects to a charged massless scalar and
fermionic fields, respectively, on a idealized cosmic string spacetime. There a 
magnetic field running through the line singularity was considered. As a 
consequence, the renormalized vacuum expectation value associated with the
energy-moment tensor, $\langle T_{\mu\nu}(x)\rangle_{Ren.}$, presents contributions
coming from the geometry of the spacetime and also the magnetic flux. In order 
to develop these calculations the respective Green functions in this physical
system were obtained. More recently Sriramkumar \cite{Sri} has calculated the
vacuum fluctuation of current and energy densities for a massless charged scalar
field around an idealized cosmic string carrying a magnetic flux. There the
Green function was obtained taking into account the presence of the vector
potential in the differential operator, $D_\mu=\partial_\mu-ieA_\mu$, which
presents the advantage of calculating the two-points function without imposing any
boundary condition on the filed. 

The analysis of the vacuum polarization effects on a massless charged 
scalar field by a magnetic flux confined in a cylindrical tube of finite radius 
in a cosmic string background have been analysed in \cite{Spinelly}. There three 
distinct configurations of magnetic field below specified has been considered.
In all of them the axis of the infinitely long tube of radius $R$ 
coincides with the cosmic string. Calculating the renormalized vacuum expectation 
value (VEV) of the square of the field, $\langle\Phi^*(x)\Phi(x)\rangle_{Ren}$,  
and the energy-momentum tensor, $\langle T^\nu_\mu(x)\rangle_{Ren}$, it was 
observed that these quantities present two contributions for each model of magnetic 
flux. The first are the standard ones due to the conical geometry of the spacetime 
and the magnetic flux. The second contributions are corrections due to the finite 
thickness of the radius of the tube. These extra terms provided relevant 
contributions for points outside the tube. Specifically, for the third model this 
contribution is a long-range effect, i.e., it is as relevant as the standard one up 
to a distance which exceed the radius  of the observable universe. 

Here in this paper we decided to continue this analysis, extending it to 
charged massless fermionic fields in the same physical system. As we shall see,
the renormalized VEV of the zero-zero component of the energy-momentum tensor 
also present two distinct contributions. Moreover due mainly to interaction of the 
spin degrees of freedom of the fermionic field with the magnetic flux, the 
corrections due to the nonzero thickness of the radius of the tube is composed
by two terms with opposite signal. Additional couplings between the
spin with the geometry are also present in this system.

The idealized model for a infinitely long straight static cosmic string 
spacetime can be given in cylindrical coordinates by the line element below:
\begin{equation}
ds^2=-dt^2+dr^2+\alpha^2 r^2d\theta^2+dz^2 \ , 
\label{1}
\end{equation}
where $\alpha=1-4\mu$ is a parameter smaller than unity which codifies the 
presence of a conical two-surface $\left( r,\theta \right)$. In fact for a 
typical Grand Unified Theory, $\alpha=1-O(10^{-6})$.

We shall consider the presence of a magnetic field along the $z$-direction 
assuming that the field has a finite range in the radial coordinate. We are
particularly interested in the three models below:\\ 
i) $H(r)=\frac{\phi}{\alpha\pi R^2}\Theta(R-r)$, homogeneous field inside;\\ 
ii) $H(r)=\frac{\phi}{2\pi\alpha Rr}\Theta(R-r)$, field proportional to $1/r$;\\ 
iii) $H(r)=\frac{\phi}{2\pi\alpha R}\delta(r-R)$, cylindrical shell,\\ 
where $R$ is the radial extent of the tube, and $\phi$ is the total flux. The ratio of the flux by the quantum flux $\phi_{o}$, can be expressed by 
$\phi/\phi_0=N+\gamma$, where $N$ is the integer part and $0<\gamma<1.$\footnote{In this paper we are considering $\hbar=G=c=1$.}

\section{SPINOR FEYNMAN PROPAGATOR}

The Feynmann propagator associated with a charged fermionic field, $S_F(x,x')$, 
obeys the following differential equation:
\begin{equation}
\left(i\n +e\A -M\right)S_F(x,x')=\frac1{\sqrt{-g}}
\delta^4(x,x')I_4 \ ,
\end{equation}
where $g=det(-g_{\mu\nu})$. The covariant derivative operator reads
\begin{equation}
\n=e^\mu_{(a)}\gamma^{(a)}(\partial_\mu+\Gamma_\mu) \ ,
\end{equation}
$e^\mu_{(a)}$ being the vierbein satisfying the condition $e^\mu_{(a)}e^\nu_{(b)}
\eta^{(a)(b)}=g^{\mu\nu}$ and $\Gamma_\mu$ is the spin connection given in terms
of flat spacetime $\gamma$ matrices by
\begin{equation}
\Gamma_\mu=-\frac14\gamma^{(a)}\gamma^{(b)}e^\nu_{(a)}e_{(b)\nu;\mu} \ ,
\end{equation}
and 
\begin{equation}
\A=e^\mu_{(a)}\gamma^{(a)}A_\mu \ .
\end{equation}

It can be shown that if a bispinor $D_F(x,x')$ satisfies the differential equation
\begin{eqnarray}
\label{D}
\left[-{\cal{D}}^2+\frac14R+ieg^{\mu\nu}(D_\mu A\nu)-ie\Sigma^{\mu\nu}F_{\mu\nu}+\right.\nonumber\\
2ieg^{\mu\nu}A_\nu\nabla_\nu
\left.+e^2g^{\mu\nu}A_\mu A_\nu+M^2\right]D_F(x,x')\nonumber\\
=-\frac1{\sqrt{-g}}\delta^4(x,x')I_4 \ ,
\end{eqnarray}
with
\begin{equation}
\Sigma^{\mu\nu}=\frac14[\gamma^\mu,\gamma^\nu] \ ,
\end{equation}
being the generalized d'Alembertian operator given by
\begin{eqnarray}
{\cal{D}}^2=g^{\mu\nu}\nabla_\mu\nabla_\nu=g^{\mu\nu}\left(\partial_\mu\nabla_\nu
+\Gamma_\mu\nabla_\nu-\Gamma^\alpha_{\mu\nu}\partial_\alpha\right) \ ,\nonumber
\end{eqnarray}
then the spinor Feynman propagator may be written as\footnote{In the absence of
gauge field this result can be found in\cite{Davies}.}
\begin{equation}
S_F(x,x')=\left(i\n+e\A+M\right)D_F(x,x') \ .
\end{equation}

Now after this brief review about the calculation of spinor Feynmann propagator,
let us specialize it to the cosmic string spacetime in the presence
of a magnetic field along the $z-$direction. Wee shall chose the following
base tetrad:
\begin{equation}
e^\mu_{(a)} = \left( \begin{array}{cccc}
  1& 0 & 0 & 0 \\
  0 & \cos\theta & -1/\alpha r \sin\theta
 & 0\\ 
0 & \sin\theta & 1/\alpha r\cos\theta
 & 0\\ 
0 & 0 & 0& 1
                      \end{array}
               \right) \ .
\end{equation}
As to the four vector potential we have
\begin{equation}
A_{\mu}=(0,0,A(r),0) \ ,
\end{equation} 
with
\begin{equation}
A(r)=\frac{\phi}{2\pi}a(r) \ . 
\end{equation}

For the two first models, we can represent the radial function $a(r)$ by:
\begin{equation}
\label{a1}
a(r)=f(r)\Theta (R-r)+\Theta (r-R) \ ,
\end{equation}
with
\begin{eqnarray}
\label{a2}
f(r)=\left\{\begin{array}{cc}
r^2/R^2,&\mbox{for the model ({\it{i}}) and}\\
r/R,&\mbox{for the model ({\it{ii}}).}
\end{array}
\right.
\end{eqnarray}
For the third model,
\begin{equation}
\label{a3}
a(r)=\Theta(R-r).
\end{equation}
The only non-zero spin connection is
\begin{equation}
\Gamma_2=\frac i2(1-\alpha)\Sigma^3
\end{equation}
and for the Christofell symbols we have:
\begin{equation}
\Gamma_{22}^1=-\alpha^2 r \ , \ \ \Gamma_{12}^2=\Gamma_{21}^1=1/r \ .
\end{equation}

Defining by ${\cal{K}}(x)$ the $4\times 4$ matrix differential operator which acts
on $D_F(x,x')$ in (\ref{D}), for this physical system we obtain:
\begin{eqnarray}
{\cal{K}}(x)&=&-{\Delta}-\frac i{\alpha^2 r^2}(1-\alpha)\Sigma^3\partial_\theta
-eH(r)\Sigma^3\nonumber\\
&+&\frac1{4\alpha^2 r^2}(1-\alpha)^2-\frac e{\alpha^2 r^2}(1-\alpha)A(r)\Sigma^3\nonumber\\
&+&\frac{2ie}{\alpha^2 r^2}A(r)\partial_\theta+\frac{e^2}{\alpha^2 r^2}A^2(r)+M^2\ , \nonumber\\
\end{eqnarray}
where
\begin{equation}
\Sigma^3=\left( \begin{array}{cccc}
  \sigma^3& 0\\ 
0 & \sigma^3
                      \end{array}
               \right) \ ,
\end{equation}
and 
\begin{equation}
{{\Delta}}=-\partial_t^2+\partial_r^2+\frac 1r\partial_r+\frac 1{\alpha^2 r^2}
\partial^2_\theta+\partial_z^2 \ .
\end{equation}

We can see that the differential operator above explicitly  exhibits, besides the
ordinary d'Alembertian operator on the conical spacetime, four different types of
interactions terms: $(i)$ the usual charge-magnetic field, $(ii)$ the 
spin-magnetic field, $(iii)$ the spin-geometry and $(iv)$ the spin-charge-geometry.
All of the last three interactions were absent in the analogous differential 
operator used to define the scalar Green function in \cite{Spinelly}.

Moreover as we can see the operator ${\cal{K}}(x)$ is diagonal in $2\times2$ blocks.
This means that the two upper components of the Dirac spinor interact with the 
gravitational and magnetic fields in similar way as the two lower components and they
do not interact among themselves.

The system that we what to study consists of massless charged fermionic field in the
cosmic string spacetime in the presence of an Abelian magnetic field along the
$z-$direction. Let us chose a left-handed field. In this case the Dirac equation
and the equation which defines the spinor Feynman propagator reduce themselves to
a $2\times 2$ matrix differential equation as shown below:
\begin{equation}
\N\chi=0 \ ,
\end{equation}
where
\begin{eqnarray}
\N=i\left[\partial_t-\sigma^{(r)}\left(\partial_r-\frac{(\alpha^{-1}-1)}{2r}\right)\right.\nonumber\\
\left. -\frac1{\alpha r}\sigma^{(\theta)}\left(\partial_\theta-iA(r)\right)-\sigma^{(z)}
\partial_z\right] \ ,
\end{eqnarray}
with $\sigma^{(r)}={\vec\sigma}.{\hat{r}}$, $\sigma^{(\theta)}={\vec{\sigma}}.
{\hat{\theta}}$ and $\sigma^{(z)}={\vec\sigma}.{\hat{z}}.$

The Feynman two-components propagator obeys the equation
\begin{equation}
i\N S_F^L(x,x')=\frac1{\sqrt{-g}}\delta^4(x,x')I_2 \ ,
\end{equation}
and can be given by
\begin{equation}
S_F^L(x,x')=i\N G^L(x,x') \ ,
\end{equation}
where now $G^L(x,x')$ obeys the $2\times 2$ matrix differential operator below:
\begin{equation}
\label{K}
{\bar{K}}(x)G^L(x,x')=-\frac1{\sqrt{-g}}\delta^4(x,x')I_2 \ ,
\end{equation}
with
\begin{eqnarray}
{\bar{K}}(x)&=&-{\Delta}-\frac i{\alpha^2 r^2}(1-\alpha)\sigma^3\partial_\theta
-eH(r)\sigma^3\nonumber\\
&+&\frac1{4\alpha^2 r^2}(1-\alpha)^2-\frac e{\alpha^2 r^2}(1-\alpha)A(r)\sigma^3\nonumber\\
&+&\frac{2ie}{\alpha^2 r^2}A(r)\partial_\theta+\frac{e^2}{\alpha^2 r^2}A^2(r)\ .\nonumber\\
\end{eqnarray} 

Because of the peculiar diagonal form of the above operator, let us take for $G^L$
the following expression:
\begin{equation}
\label{GL} 
G^L(x,x') = \left( \begin{array}{cccc}
  G_+(x,x')& 0 \\ 
0 & G_-(x,x')
                      \end{array}
               \right) \ .
\end{equation}
So the differential equation (\ref{K}) reduces itself to a two independent ones as
shown below:
\begin{eqnarray}
\label{G1}
\left[{\Delta}\pm\frac i{\alpha^2 r^2}(1-\alpha)\partial_\theta-\frac{(1-\alpha)^2}
{4\alpha^2 r^2}\pm eH(r)\right.\nonumber\\
\left.-\frac{2ie}{\alpha^2 r^2}A(r)\partial_\theta \pm \frac{e}{\alpha^2 r^2}(1-\alpha)A(r)-\right.\nonumber\\
\left.\frac{e^2}{\alpha^2 r^2}A^2\right]G_{\pm}(x,x')=\frac1{\sqrt{-g}}\delta^4(x,x') \ .
\end{eqnarray}

Due to the cylindrical symmetry of this system each component of the Euclidean 
Green function can be expressed by
\begin{eqnarray}
G_{\pm}(x,x')=\frac1{(2\pi)^3}\sum_{n=-\infty}^\infty e^{in(\theta-\theta')}
\int_{-\infty}^\infty dk \nonumber\\
\int_{-\infty}^{\infty} d\omega e^{ik(z-z')}e^{i\omega (\tau-\tau')} g_n^{\pm}(r,r') \ .
\label{11}
\end{eqnarray}

Before to specialize on the specific model let us write down the non-homogeneous
differential equation obeyed by the unknown function $g_n^\pm(r,r')$.
Substituting (\ref{11}) into (\ref{G1}) and using the standard representation to 
the delta function in the temporal, angular and $z$-coordinates, we arrive to 
following differential equation for the unknown function $g_n^{\pm}(r,r')$:
\begin{eqnarray}
\left[\frac{d^2}{dr^2}+\frac{1}{r}\frac d{dr}-\frac1{\alpha^2r^2}\left[n^2\pm n(1
-\alpha \mp 2neA)\right. \right. \nonumber\\
+\left.\left.\frac{(1-\alpha^2)}4 \mp e(1-\alpha)A(r)+e^2A^2\right] \right. \nonumber\\
\left. -\beta^2 \pm H(r) \right]g_n^\pm(r,r')=\frac1{\alpha r}\delta(r-r') 
\label{12}
\end{eqnarray}
where $\beta^2=k^2+\omega^2$.

It is of our main interest to investigate the vacuum polarization effect for 
external points to the magnetic flux. So, we shall consider solutions of (\ref{12}) 
with both $r$ and $r'$ greater than $R$.

Let us first consider the models ({\it{i}}) and ({\it{ii}}). The inner 
solution of (\ref{12}), corresponding to $r<r'$ is $g^<_n(r,r')$. Integrating 
out in region $r<R$, we have:
\begin{equation}
g^{\pm <}_n(r,r')=A_{(i)}^\pm H_{i}^\pm(r)\ ,
\label{13}
\end{equation}
for $r<R$ and
\begin{eqnarray}
g^{\pm <}_n(r,r')&=&B_{(i)}^\pm\left[I_{|\nu_{\pm}|}(\beta r)\right. \nonumber\\
&&\left. +E_{(i)}^\pm(\beta R)K_{|\nu_{\pm}|}(\beta r)\right],
\label{14}
\end{eqnarray}
for $R<r<r^{'}$, where $\nu_{\pm}=\frac{(n \pm \frac{1-\alpha}2-\delta)}\alpha$, being $\delta=
\frac{e\phi}{2\pi}=N+\gamma$. $H_{i}^\pm(r)$, for $i=1,2$, represents the solution 
associated with the two first models:
\begin{equation}
H_{1}^\pm(r)=\frac{1}{r}M_{\sigma_{1(\pm)},|\lambda_{1(\pm)}|}\left( \frac{\delta}{\alpha R^2} 
r^2\right),
\label{15}
\end{equation}
and
\begin{equation}
H_{2}^\pm(r)=\frac{1}{\sqrt{r}}M_{\sigma_{2(\pm)},|\lambda_{2(\pm)}|}\left( \zeta r\right),
\label{16}
\end{equation}
with $\sigma_{1(\pm)}=(n-\frac{\beta^{2}R^{2}\alpha^{2}}{2\delta}\pm\frac{1}{2}(\alpha+1))/2\alpha$, $\lambda_{1(\pm)}=n/2\alpha\pm(1-\alpha)/{4\alpha}$, $\sigma_{2(\pm)}=\frac{\delta(2n\pm 1)}{2\alpha(\delta^{2}+\beta^{2}\alpha^{2}R^{2})^{1/2}}$ and $\lambda_{2(\pm)}=n/\alpha\pm(1-\alpha)/{2\alpha}$.
For both functions $H_2^\pm$, $\zeta=\frac{2}{R\alpha}(\delta^2+\beta^2 R^2 
\alpha^2)^{1/2}$. 

In both cases $M_{\sigma,\lambda}$  represents the Whittaker function. The constant 
$E_{i}^\pm(\beta R)$ is determined by matching $g_{n}^{\pm<}$ 
and its first derivative at $r=R$. So we get 
\begin{equation}
\label{E1}
E_i^\pm(\beta R)=\frac{S_i^\pm(\beta R)}{P_i^\pm(\beta R)} \ ,
\label{17}
\end{equation} 
where
\begin{eqnarray}
S_i^\pm(\beta R)=H_i'^\pm(R)I_{|\nu_\pm|}(\beta R)-H_i^\pm(R)I'_{|\nu_\pm|}\nonumber
(\beta R)
\end{eqnarray}
and
\begin{eqnarray}
P_i^\pm(\beta R)=H_{i}^\pm(R)K'_{|\nu_\pm|}(\beta R)-H_i'^\pm(R)K_{|\nu_\pm|}(\beta R).\nonumber
\end{eqnarray}
In all the expressions, $I_\nu$ and $K_\nu$ are the modified Bessel 
functions.

The outer solution of (\ref{12}) is given by
\begin{equation}
g^{\pm>}_{n(i)}(r,r')=D_{(i)}^\pm K_{|\nu_\pm|}(\beta r), 
\quad \hbox{for $r>r'$.}
\label{18} 
\end{equation}

Now imposing the boundary conditions on $g^{\pm<}_n$ and $g^{\pm>}_n$ at $r=r'$,
we get the following result:
\begin{eqnarray}
\label{gn}
g_n^\pm(r,r')=-\frac1\alpha\left[I_{|\nu\pm|}(\beta r^<)+\right.\nonumber\\
\left.E^\pm(\beta R)K_{|\nu\pm|}(\beta r^<)\right]K_{|\nu\pm|}(\beta r^>) \ .
\label{19}
\end{eqnarray} 
Substituting (\ref{gn}) into (\ref{11}) and after some intermediate
steps, the Euclidean Green function acquires the following expression:
\begin{eqnarray}
\label{Ga}
G_\pm(x,x')=-\frac{e^{i N\Delta\theta}}{8\pi^2\alpha rr'\sinh u_0}\times &&\nonumber\\
\frac{e^{\mp i\Delta \theta}\sinh(\gamma^\pm u_0/\alpha)+\sinh[(1-\gamma^\pm)u_0/
\alpha]}{\cosh(u_0/\alpha)-\cos\Delta\theta}
\nonumber\\
-\frac{1}{4 \pi^{2}}\int_0^{\infty} d\beta \beta J_{0}\left(\beta\sqrt{(\Delta 
\tau)^{2}+(\Delta z)^{2}}\right)\times
\nonumber\\
\sum_ne^{i n\Delta\theta}E^\pm_n(\beta R)K_{|\nu_\pm|}(\beta r)K_{|\nu_\pm|}
(\beta r') \ ,
\label{19}
\end{eqnarray}
with $\gamma^\pm=(1-\alpha)/2 \pm \gamma$ and
\begin{equation}
\cosh u_{o}=\frac{r^2+{r^{'}}^{2}+(\Delta \tau)^{2}+(\Delta z)^{2}}{2rr^{'}} \ .
\label{20}
\end{equation}

As to the third model, the solution to $g_{n}^\pm(r,r')$ is:
\begin{equation}
g^{\pm<}_n(r,r')=A^\pm I_{|\nu_\pm|}(\beta r) \ ,
\end{equation}
for $r<R$,
\begin{eqnarray}
g^{\pm<}_n(r,r')&=&B^\pm\left[I_{|\epsilon_\pm|}(\beta r)\right. \nonumber\\
&&\left.+E^\pm(\beta R)K_{|\epsilon_\pm|}(\beta r)\right] \ ,
\end{eqnarray}
for $R<r<r^{'}$ and
\begin{equation}
g^{\pm>}_n(r,r')=D^\pm K_{|\epsilon_\pm|}(\beta r), 
\label{22} 
\end{equation}
for $r>r^{'}$. In the above equation $\epsilon\pm=1/\alpha(n\pm(1-\alpha)/2)$.
Again the coefficient $A^\pm$, $B^\pm$ and $D^\pm$ can be determined by imposing
boundary conditions at $r=R$ and $r=r'$. However, due the expression
to the magnetic field in this model is concentrate as a $\delta-$function
at the cylindrical shell, there happens a discontinuity condition in the
first derivative of $g^{\pm<}$ at $r=R$. The rest being the same. So
using these facts we obtain:
\begin{eqnarray}
g^{(\pm)}_n(r,r')&=&\frac1\alpha\left[E^\pm(\beta R)K_{|\epsilon_\pm|}(\beta r^<)\right. \nonumber\\
&&\left. +I_{|\epsilon_\pm|}(\beta r^<)\right]K_{|\epsilon_\pm|}(\beta r^>) \ ,
\end{eqnarray}
where now
\begin{equation}
E^\pm(\beta R)=\frac{S^\pm(\beta R)}{P^\pm(\beta R)}
\label{E2}
\end{equation}
with
\begin{eqnarray}
S^\pm(\beta R)=I_{|\epsilon_\pm|}(\beta R)I'_{|\nu_\pm|}(\beta R)-&&\nonumber\\
I_{|\nu_\pm|}(\beta R)
I^{'}_{|\epsilon_\pm|/\alpha}(\beta R) \mp\frac{\delta}{\alpha R}I_{|\epsilon_\pm|}(\beta R)I_{|\nu_\pm|}(\beta R)&& \nonumber
\end{eqnarray}
and
\begin{eqnarray}
P^\pm(\beta R)=I_{|\nu_\pm|}(\beta R)K^{'}_{\epsilon_\pm}(\beta R)-&&\nonumber\\
I_{|\nu_\pm|}^{'}(\beta R)K_{|\epsilon_\pm|}(\beta R)\pm \frac{\delta}{\alpha R} K_{|\epsilon_\pm|}(\beta R)I_{|\nu_\pm|}(\beta R).&& \nonumber
\end{eqnarray}

Finally substituting the expression found to $g^\pm_n$ above into 
(\ref{11}), and adopting similar procedure as we did in the two previous 
cases, we obtain:
\begin{eqnarray}
\label{Gb}
&&G_\pm(x,x')=-\frac1{8\pi^2 \alpha rr'\sinh u_0}\times \nonumber\\
&&\frac{e^{\mp\Delta\theta}
\sinh({\bar{\gamma}}u_0/\alpha)+\sinh[(1-{\bar{\gamma}})u_0/\alpha]}
{\cosh(u_{o}/\alpha)-\cos\Delta\theta}
\nonumber\\
&-&\frac1{4 \pi^{2}}\int_0^{\infty} d\beta \beta J_{0}\left(\beta\sqrt{(\Delta 
\tau)^{2}+(\Delta z)^{2}}\right)\times
\nonumber\\
&&\sum_{n=-\infty}^\infty e^{in\Delta\theta}
E^\pm(\beta R)K_{|\epsilon_\pm|}(\beta r)K_{|\epsilon_\pm|}(\beta r') \ ,\nonumber\\
\label{24}
\end{eqnarray}
with ${\bar{\gamma}}=(1-\alpha)/2$.

\section{COMPUTATION OF $\langle\hat{T}_{00}\rangle_{Ren}$}

The vacuum expectation value of the energy-momentum 
tensor associated with this system under investigation is given by:
\begin{eqnarray}
\langle\hat{T}_{\mu\nu}\rangle&=&\frac14\lim_{x'\to x}tr\left[\sigma_\mu(D_\nu-{\bar{D}}_{\nu'})\right. \nonumber\\
&&\left. -\sigma_\nu(D_\mu-{\bar{D}}_{\mu'})\right]S_F^L(x,x') \ , \label{45}
\end{eqnarray}
where $D_\sigma=\nabla_\sigma-ieA_\sigma$ and the bar denotes complex 
conjugate and $\sigma^\mu=(I_2,\sigma^{(r)},\sigma^{(\theta)},\sigma^{(z)})$ . 

In order to take into account the presence of the three magnetic field 
configurations given previously we write the vector potential in the form 
$A_\mu=(0,0,\frac{\pi a(r)}{2\pi\alpha},0)$, with $a(r)$ being given by 
(\ref{a1}) and (\ref{a2}), for the first two cases and by (\ref{a3}) for the third 
case. The spinor Green functions are expressed in terms of the bispinor $G^L(x,x')$ 
given by (\ref{GL}) with $G_\pm(x,x')$ given by (\ref{Ga}) and (\ref{Gb}).

For simplicity let us calculate $\langle T_{00}(x)\rangle$ only. The other 
components of the VEV of the energy-momentum tensor can be obtained by using 
the conservation condition and trace anomaly. 
Fortunately only the second order time derivative provides a nonzero contribution 
to $\langle T_{00}(x)\rangle$. In fact all the other terms go to zero in the
coincidence limit and/or after taking the trace over the Pauli matrices. 
Moreover, because the bispinor depends on the time variable with $t-t'$, we
finally have:
\begin{eqnarray}
\langle T_{00}(x)\rangle&=&-\lim_{x'\to x}tr(\partial_t^2G(x,x'))\nonumber\\
&=&\lim_{x'\to x}tr(\partial_\tau^2G_E(x,x')) \ ,
\end{eqnarray}
where we have made a Wick rotation on the above equation.

However the calculation of the above expression provides a divergent 
result. In order to obtain a finite and well defined result, we must
apply in this calculation some renormalization procedure. Here we
shall adopt the following prescription: we subtract from the Green 
function the Hadamard one before applying the coincidence limit as 
shown:
\begin{eqnarray}
\langle T_{00}(x)\rangle_{Ren.}&=&\lim_{x'\to x}tr[\partial_\tau^2G_E(x,x')\nonumber\\
&&-\partial_\tau^2G_H(x,x')] \ .
\end{eqnarray}
Because this spacetime is locally flat, the Hadamard function coinciding with the 
Euclidean Green function in a flat spacetime: $G_H(x,x')=
\frac1{4\pi}\frac1{(x-x')^2}I_2$. As it was already mentioned, (\ref{Ga}) and
(\ref{Gb}), present two distinct contributions; the first ones contain informations
about the geometrical structure of the spacetime and the fractional part of the
magnetic flux, and the second, the corrections, are due to the nonzero thickness
of the radius of tube. Moreover, in the calculation of the VEV, only their first
contributions are divergent in the coincidence limit, the second are finite. Finally,
explicitly exhibiting these remarks, we write down the renormalized VEV of the
zero-zero component of the energy-moment tensor by:
\begin{equation}
\langle T_{00}(x)\rangle_{Ren.}=\langle T_{00}(x)\rangle_{Reg.}+
\langle T_{00}(x)\rangle_{C} \ ,
\end{equation}
where
\begin{eqnarray}
\langle T_{00}(x)\rangle_{Reg.}&=&\lim_{x'\to x}\left[\partial_\tau^2G_+(x,x')
\right. \nonumber\\ 
&&\left. +\partial_\tau^2G_-(x,x')-2\partial_\tau^2G_H(x,x')\right]\nonumber\\
\end{eqnarray}
and 
\begin{equation}
\langle T_{00}(x)\rangle_{C}=\lim_{x'\to x}\partial_\tau^2G_C(x,x') \ ,
\end{equation}
where, $G_C(x,x')$, represents the corrections due to the second terms in $G_+$ and
$G_-$ for the three models, as shown next. After some intermediate calculations we 
arrive to the following results below:\\
\noindent
$i)$ For the two first models,
\begin{eqnarray}
\label{T1} 
&&\langle T_{00}(x)\rangle_{Ren.}=\frac1{5760\pi\alpha^4 r^4}\times \nonumber\\ &&\left[(\alpha^2-
1)(17\alpha^2+7)+120\gamma^2(\alpha^2-2\gamma^2-1)\right]\nonumber\\
&&+\frac1{4\pi^2r^4}\int_0^\infty dv v^3\sum_{n=-\infty}^\infty\left[
E_i^+(vR/r)K_{|\nu_+|}^2(v)\right. \nonumber\\
&&\left.+E_i^-(vR/r)K_{|\nu_-|}^2(v)\right] \ ,
\end{eqnarray}
for $i=1$ and $2$. \\
\noindent
$ii)$ For the third model,
\begin{eqnarray}
\label{T2} 
&&\langle T_{00}(x)\rangle_{Ren.}=\frac1{5760\pi\alpha^4 r^4}(\alpha^2-
1)(17\alpha^2+7)\nonumber\\
&&+\frac1{4\pi^2r^4}\int_0^\infty dv v^3\sum_{n=-\infty}^\infty\left[E^+(vR/r)
K_{|\epsilon+|}^2(v)\right. \nonumber\\
&&\left. +E^-(vR/r)K_{|\epsilon-|}^2(v)\right] \ .
\end{eqnarray} 
For all the above expressions the coefficient $E^\pm$ were given in (\ref{E1})
and (\ref{E2}). 

Before to provide some qualitative information about the second contributions 
of (\ref{T1}) and (\ref{T2}), we would like to make a few comments about our
results: $(i)$ The first contributions of them depend only on the conicity
parameter $\alpha$, and the fractional part of $\phi/\phi_0$, denoted by 
$\gamma$, for the two first models. $(ii)$ The second contributions, the
corrections, vanish in the limit $R\to 0$. They also depend on the integer
part of $\phi/\phi_{0}$, $N$.

Some qualitative information about the behavior of the second contributions of
(\ref{T1}) and (\ref{T2}) can be provided. Although for the three models these
contributions present an overall $1/r^4$ dependence on the radial coordinate,
there exist an additional dependece in their integrand by the coefficient $E^\pm$. 
By our numerical analysis each term present opposite signal; however they do
not cancel each other. Our analysis were developed for $N=0$ and $N=1$, for
$\alpha=0.99$, $\gamma=0.02$ and $R/r=10^{-3}$.

\section{CONCLUDING REMARKS}

In this paper we have explicitly exhibited the spinor Green functions associated
with a charged left-handed field on a cosmic string spacetime in the presence of
external magnetic fluxes in three different configurations, all of them confined in 
a long tube of radius $R$. Having these Green functions we calculated the
formal expression to the renormalized VEV of the zero-zero component of 
energy-momentum tensor, $\langle T_{00}(x)\rangle_{Ren}$. In these calculations we 
observe that two independent contributions were obtained. The first contributions 
are the standard ones due to the conical geometry of the spacetime and the magnetic 
flux. They coincide with the expression found by Linet in \cite{Linet}. The second 
contributions are corrections due to the finite thickness of the radius of the tube.
Unfortunately, it was not possible to obtain an analytical expression to provide
the dependence of these second contributions with the radial coordinate, and 
only by numerical analysis we can do this. However some qualitative information 
can be given: we found for specific values of the parameter that the integrands
of these corrections have opposite signal, however they do not cancel each other.
Our next task will be to develop a more systematic numerical analysis to provide
quantitative information about the second parts of  (\ref{T1}) and (\ref{T2}).

{\bf{Acknowledgments}}
\\       \\
We would like to thank Conselho Nacional de Desenvolvimento Cient\'\i fico e 
Tecnol\'ogico (CNPq.) and CAPES for partial financial support.

\end{document}